\documentstyle[emulateapj,10pt,apjfonts,psfig]{article}

\def\snr{B0453--685}
\def\cxo{{\em Chandra}}

\def\kms{km~s$^{-1}$}
\def\etal{{\rm et~al.\ }}

\lefthead{GAENSLER ET AL.}
\righthead{PULSAR WIND NEBULA IN THE LMC}

\begin{document}
\title{Discovery of a New Pulsar Wind Nebula in the Large Magellanic Cloud}
\author{B. M. Gaensler\altaffilmark{1}, 
S. P. Hendrick\altaffilmark{2},
S. P. Reynolds\altaffilmark{1, 2} and K. J. Borkowski\altaffilmark{2}}
\altaffiltext{1}{Harvard-Smithsonian
Center for Astrophysics, 60 Garden Street MS-6, Cambridge, MA 02138;
bgaensler@cfa.harvard.edu}
\altaffiltext{2}{Department of Physics, North Carolina State University, Raleigh, NC
27695}

\begin{abstract}

We present new high-resolution radio and X-ray observations of the
supernova remnant (SNR) \snr\ in the Large Magellanic Cloud, carried out
with the Australia Telescope Compact Array and the {\em Chandra X-ray
Observatory}\ respectively. Embedded in the SNR shell is a compact
central nebula producing both flat-spectrum polarized radio emission
and non-thermal X-rays; we identify this source as a pulsar wind nebula (PWN)
powered by an unseen central neutron star.
We present a new approach by which 
the properties of a SNR and PWN can be used to infer
upper limits on the initial spin period and surface magnetic
field of the unseen pulsar, and conclude that this star was an initial
rapid rotator with current properties similar
to those of the Vela pulsar.  As is the case for other similarly-aged
sources, there is currently an interaction taking place between the PWN
and the SNR's reverse shock.

\end{abstract}


\section{Introduction}

The Magellanic Clouds are ideal regions for studying supernova remnants
(SNRs), the known distance and low extinction
obviating many of the frustrations associated with studying Galactic
SNRs. While this has proved useful in studying the shell emission
produced by the interaction of SNRs with the ambient medium,
just two Magellanic SNRs, B0540--693 (\cite{msk93}) and N157B
(\cite{wgcd01}), are known to harbor a young energetic pulsar and its
surrounding pulsar wind nebula (PWN). The lack of other PWNe in the
Magellanic Clouds is likely due to a lack of spatial resolution
--- such resolution is only now being applied to Magellanic SNRs in both
the radio and X-ray bands, and so the prospects are good for identifying
new ``composite'' SNRs, containing both outer shells and central PWNe.
Such sources are of particular interest, as the shell and PWN together
provide unique constraints on the properties of the system.

We here present data which demonstrate that \snr\ in the 
LMC is a composite SNR.  This {\em Letter}\ primarily discusses
the central PWN; interpretation of the
surrounding shell will be the focus of a subsequent paper (\cite{hen03}).

\section{Observations and Results}

Radio observations of \snr\ were carried out with the Australia
Telescope Compact Array (ATCA). Observations were made simultaneously
at 1.4 and 2.4~GHz with a bandwidth of 128~MHz at each frequency,
using the 1.5G\footnote{1.5G is a non-standard ATCA configuration, and
has successive spacings between the six antennas of 30.6, 535.7, 306.1,
413.3 and 3015.3~meters.} and 6.0C configurations on 2002 Jul 23
and 2002 Aug 26 respectively. A 38.6-ks X-ray observation of the same field
was carried out using the ACIS-S3 detector of the {\em
Chandra X-ray Observatory}\ on 2001~Dec~18.

For each radio observation, the time on source was $\sim$11~hr.
Antenna gains and polarization leakages were calibrated using
observations of PKS~B0252--712, while the flux density scale was tied
to PKS~B1934--638. The radio data were edited, calibrated and
imaged using standard techniques;\footnote{See {\tt
http://www.atnf.csiro.au/computing/software/miriad} .}  the resulting resolution
at each frequency is listed in Table~\ref{tab_src}.\footnote{The 1.4-GHz
image was formed using all antennas, while the 2.4-GHz image excluded
the 6th antenna so as to not overresolve the source.}
Analysis of the \cxo\ data will be reported elsewhere
(\cite{hen03}).

\subsection{Imaging and Polarimetry}

\begin{figure}[hbt]
\centerline{\psfig{file=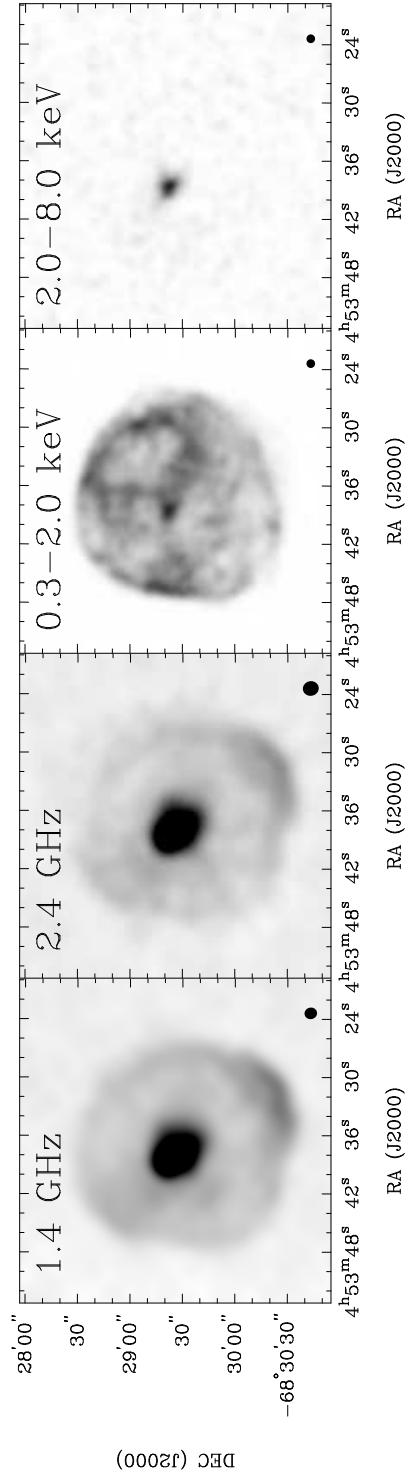,height=0.8\textheight}}
\caption{Radio and X-ray images of SNR~\snr. The top two panels show ATCA
radio images of the source at 1.4 and 2.4~GHz, while the
bottom two panels show \cxo\ X-ray images of \snr\ in the soft and hard
bands. The two radio images have both been smoothed to
a resolution of $10''\times10''$.  The sensitivity of the radio images
are 95 and 65~$\mu$Jy~beam$^{-1}$ at 1.4 and 2.4~GHz respectively.
Both radio images are shown over the same greyscale range, --0.4 to
+5.7~mJy~beam$^{-1}$.  The X-ray images have both been smoothed with
a gaussian of FWHM $5''\times5''$; the greyscale ranges are
0\%--100\% and 0\%--56\% of the peak for the soft- and hard-band
images respectively.
In all four panels, the 
point-spread function is shown by the ellipse in the
lower-right corner, and the greyscale is shown using a
linear transfer function.}
\label{fig_all_bands}
\end{figure}

Radio and X-ray images of SNR~\snr\ are shown in
Figure~\ref{fig_all_bands}.  At radio frequencies, the source is clearly
resolved into two components: a bright central core, embedded in
an approximately circular limb-brightened shell. Table~\ref{tab_src}
lists the flux densities of these two components at 1.4 and 2.4~GHz after
applying a correction for the local background. 
Fitting an ellipse to the perimeter of the
shell component, we find that if one excludes the bright protrusion to
the southwest, the shell can be fitted by a circle of diameter $120''\pm4''$,
centered at RA (J2000) $04^{\rm h}53^{\rm m}37\fs2$, Decl.\ (J2000)
$-68^\circ 29'30''$ (with an uncertainty $\pm2''$).
For a distance $50d_0$~kpc,
the diameter of the shell is $(29\pm1)d_0$~pc.

\begin{figure}[hbt]
\centerline{\psfig{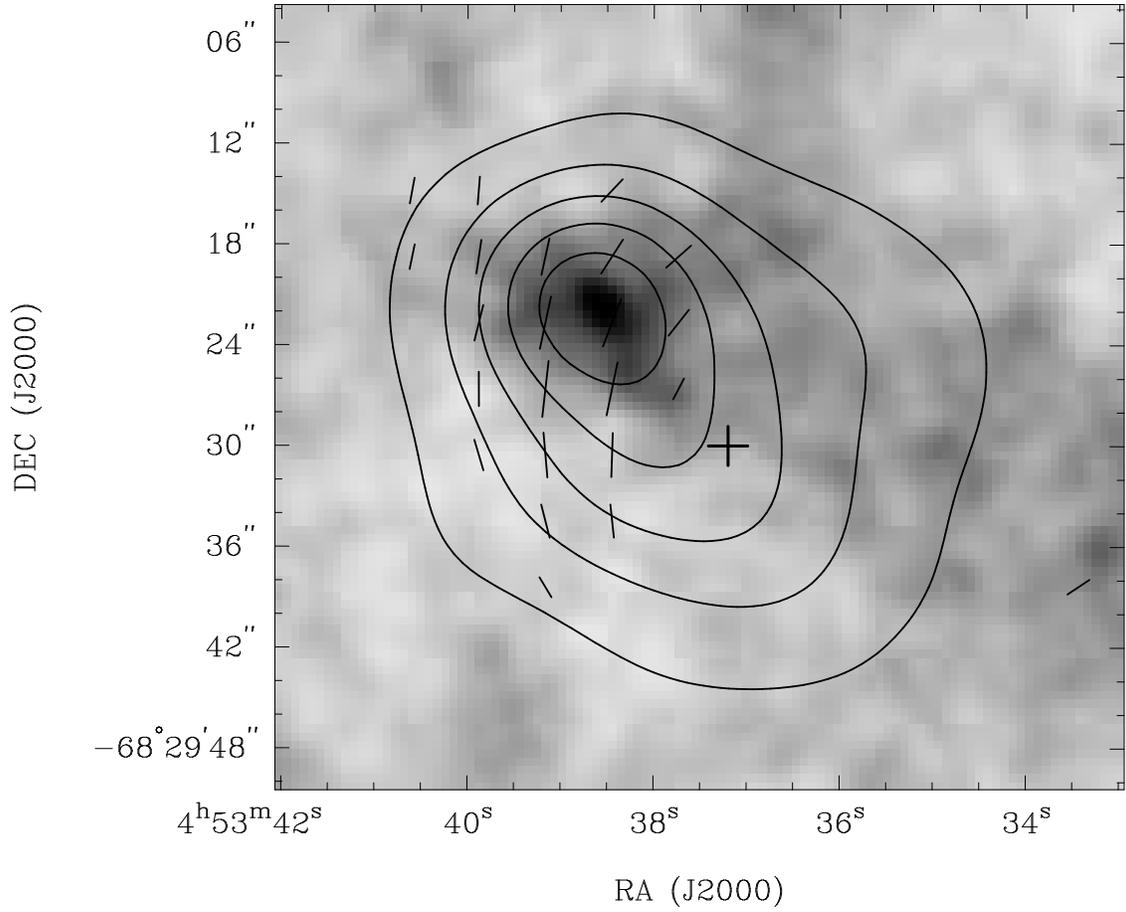}}
\caption{Radio and X-ray images of the core of SNR~\snr. The greyscale
shows X-ray emission from the central region of \snr\ as seen by \cxo\
in the energy range 0.3--8.0~keV, smoothed to a resolution of $2''$. The
contours represent the 2.4-GHz image at full resolution ($9\farcs2 \times
8\farcs4$),
drawn at the levels of 2, 4, 6, 8 and 10~mJy~beam$^{-1}$; the rms
sensitivity of these data are 70~$\mu$Jy~beam$^{-1}$.
Overlaid as vectors are the 2.4-GHz linearly polarized intensity at each
position; the length of each vector is proportional to the polarized
intensity (up to a maximum of 0.7~mJy~beam$^{-1}$) and the orientation
of the vector indicates the mean position angle of the electric
field (averaged across the observing bandwidth, and not corrected for
Faraday rotation). The ``+'' symbol indicates the center fitted for the
surrounding SNR shell.}
\label{fig_core}
\end{figure}

Figure~\ref{fig_core} shows that the central radio core is
elongated along a position angle $\sim45^\circ$, with 
dimensions $30''\times20''$ (i.e.\ a spatial size
$7.3d_0$~pc $\times$ $4.8 d_0$~pc).  The core is brightest in the
NE, where it is dominated by a
slightly extended region, with peak position (J2000) RA 
$04^{\rm h}53^{\rm m}38\fs6$, Dec.\ $-68^\circ29'21\farcs7$
($\pm0\farcs5$ in each coordinate). The core
shows significant linear polarization, of mean fractional intensity 6\%
at 1.4~GHz and 8\% at 2.4~GHz, as shown in 
Figure~\ref{fig_core}. The brightest region of the core
is offset from the center of the shell by $11''\pm2''$. To the SW,
the surface brightness of the core fades  with increasing
distance from the peak.

The third panel of Figure~\ref{fig_all_bands} demonstrates that
the morphology of \snr\ in the soft X-ray band shows a broadly similar
structure to that seen in the radio, again revealing a central core
within a limb-brightened shell.
In the hard X-ray band (lower
panel of Fig~\ref{fig_all_bands}), only the central core is visible.
The extent of this region is $14''\times7''$ ($3.4d_0$~pc $\times$
$1.7d_0$~pc), less than half that
seen in the radio. 
Figure~\ref{fig_core} demonstrates that the 
positions of the peak emission in the X-ray and
radio cores coincide to within the resolution of the observations.

\subsection{Spectral Analysis}

We have carried out a spectral index study of the radio data by smoothing
the 1.4 and 2.4~GHz images to a common resolution of $15''$ and then
applying spectral tomography to these maps (\cite{kr97}). We
find that the central core has a flatter radio spectrum than
the surrounding shell: tomography implies $\alpha =
-0.10\pm0.05$ for the core and $\alpha = -0.5\pm0.1$ for the shell
($S_\nu \propto \nu^{\alpha}$).  Assuming $\alpha =
-0.10$ for the core between $10^7$ and $10^{11}$~Hz, the radio luminosity
of this component is $L_R \sim 2d_0^2\times10^{34}$~ergs~s$^{-1}$.

At X-ray energies, we have determined the spectral properties of the
different components of the SNR by extracting events from
appropriate source and background regions and fitting to the
resulting spectra, as will be described in detail by
Hendrick \etal\ (2003\nocite{hen03}).  To briefly summarize their results,
the shell component is well-fitted by the absorbed thermal
spectrum expected in the Sedov phase of SNR evolution, while the
central core is best fit by an absorbed power law.\footnote{
Bremsstrahlung ($kT \sim 4$ keV) can also fit the central
core, but the absence of lines demands either absurdly low
abundances or an extremely underionized plasma (ionization timescale
$n_e t < 100$~yr~cm$^{-3}$).} The best-fit
parameters for the core are a foreground absorbing column
$N_H = (1.3\pm0.2)\times10^{21}$~cm$^{-2}$, a spectral index $\alpha =
-0.9\pm0.4$ and an unabsorbed flux density at 1~keV $f_X =
(6\pm1)\times10^{-14}$~erg~cm$^{-2}$~s$^{-1}$~keV$^{-1}$ 
(uncertainties quoted at 90\% confidence). 
The corresponding X-ray luminosity over the
energy range 0.5--10~keV is $L_X \sim 6d_0^2\times10^{34}$~erg~s$^{-1}$.
The spectral fits for the surrounding shell imply an age $\tau \approx 13$~kyr,
a swept-up mass $M_{sw} 
\approx280 M_\odot$, an upstream density $n_0 \approx 0.4$ cm$^{-3}$
and an initial explosion energy $E_0 \approx 5\times10^{50}$~ergs
(\cite{hen03}).

\section{Discussion}

The core embedded in \snr\ is centrally located, polarized,
has a filled-center morphology and a flat radio spectrum,
and is coincident with a smaller non-thermal X-ray nebula.
These properties demonstrate that \snr\ is a composite SNR, containing
a synchrotron-emitting PWN powered by a neutron star embedded in a
shell SNR.  

\subsection{Properties of the Central Pulsar}
\label{sec_psr}

Within the core, no radio or X-ray point source corresponding to the
central pulsar is apparent. We can calculate upper limits on the flux
from such a source as follows.  At radio wavelengths, we made images
of the field using only the longest ($>3000$~meters) five baselines in
each configuration.  At both 1.4 and 2.4~GHz, the brightest region of
the core was still detected in these images.  At the peak of nebular
emission, the corresponding upper limit on a point source is 3~mJy and
0.4~mJy at 1.4 and 2.4~GHz, respectively.

In the X-ray band, we employed the {\tt SHERPA}\ package
to carry out a 2D fit to the nebular morphology.  We fit the core with
three components: an elliptical gaussian to define the overall shape,
an unresolved source to represent an embedded pulsar, and a constant
offset to account for background.
Using these three components, we find that in the energy range 0.3--10
keV, the best-fit model contains 58 counts in an unresolved central
source.  While the apparent significance of this source seems high,
we obtained equally good or better fits by
replacing the unresolved component by a resolved Gaussian. This suggests
a complex PWN morphology and only marginal evidence for a point source.
Therefore, in subsequent discussion we adopt 58 counts as an upper limit,
corresponding to a count-rate $<1.5\times10^{-3}$~cts~sec$^{-1}$. Assuming
a column density $N_H = 1.3\times10^{21}$~cm$^{-2}$ and a power law
spectrum with a photon index $\Gamma = 1.5$ (typical for a young pulsar),
this corresponds to an unabsorbed X-ray luminosity (0.5--10~keV) $L_X <
6d_0^2 \times10^{33}$~ergs~s$^{-1}$.

These radio and X-ray upper limits on emission from a central pulsar are
not especially constraining.  Our radio limit is $\sim40$ times
above the upper limit for radio pulsations at this position
from the survey of Manchester \etal\ (2003\nocite{mfl+03}).  Several
young pulsars have been recently identified with a point-source X-ray
luminosity $L_X \approx 10^{32}-10^{33}$~ergs~s$^{-1}$ (\cite{mss+02};
\cite{hsp+03}), well below the limit seen here.

In the absence of a direct detection, a first guess
as to the properties of the pulsar in \snr\ can come from
comparison with other systems.  There are $\sim20$ other shell
SNRs known to contain a central X-ray/radio PWN; of
these other sources, the properties of \snr\ most closely
resemble those of the Vela~SNR and of G0.9+0.1, as demonstrated in
Table~\ref{tab_compare}.  The similarity of these three systems argues
that SNR~\snr\ is most likely powered by a ``Vela-like'' pulsar (e.g.\
\cite{kbm+03}), with a spin-period $P \sim 100$~ms, a surface magnetic
field $B\sim3\times10^{12}$~G and a spin-down luminosity $\dot{E}
\approx 10^{37}$~ergs~s$^{-1}$.

In cases in which a PWN has no identified pulsar, considerable
further effort has been invested to infer the pulsar's properties 
(i.e.\ current spin period $P$, initial period $P_0$, period
derivative $\dot{P}$ and surface magnetic field $B$) 
from those of its PWN. 
The age, spin-down luminosity and surface magnetic
field of the system are, respectively:
\begin{equation}
\tau =  \frac{P}{(n-1)\dot{P}} \left[ 1 - \left( \frac{P_0}{P}
\right)^{n-1}\right],~~
\dot{E} = 4\pi^2 I \frac{\dot{P}}{P^3},~~
B = 3.2\times10^{19}(P\dot{P})^{1/2}~{\rm Gauss},
\label{eqn_1}
\end{equation}
where $I\equiv10^{45}$~g~cm$^2$ is the star's moment of inertia,
$n \equiv 2- P\ddot{P}/\dot{P}^2$ is the ``braking index'',
and $P$ is in seconds.  Typically one assumes $n=3$, $P_0 = 0$ and $L_X =
\eta \dot{E}$ (where $\eta$ is an assumed efficiency factor).  
$L_X$ and $\tau$ are determined from observations, from which
estimates for $P$, $\dot{P}$ and $B$ can then be derived
using Equations~(\ref{eqn_1}).

Here we propose an alternative approach, where rather than assuming an
(unphysical) value of $P_0$, we use the above expressions to eliminate
$P$ and $\dot{P}$, and thus obtain a relation between $B$ and $P_0$.
Assuming that $B$ is constant with time, the resulting function has
absolute maxima in both $B$ and $P_0$. The robust upper limits on the
surface magnetic field and the initial spin period of the neutron star
which result are:
\begin{equation}
B < 3.2\times10^{19} \frac{U}{\tau(n-1)}~{\rm Gauss};~~~~~ 
P_0 < U (\tau[n+1])^{-1/2} \left(\frac{2}{n+1}\right)^{\frac{1}{n-1}}, 
\end{equation}
where $U^2 = P^3/\dot{P} = 4\pi^2I\eta/L_X$.  These limits are reasonably
insensitive to the value of $n$ assumed; for $2 < n < 3$
(encompassing the range for the four pulsars for which $n$ has
been directly measured; e.g.\ \cite{cmm03}), we find $B <
(3.1-6.3)\eta^{1/2}\times10^{13}$~G and $P_0 < (448-488)\eta^{1/2}$~ms
for this system.  For other PWNe, a solid
upper limit is
$\eta < 0.05$; we have argued above that the pulsar here is
``Vela-like'', for which $\eta \approx 0.01 - 0.001$ is more
reasonable (\cite{pccm02}).  We thus conclude that this
pulsar has a surface magnetic field $B <
6\times10^{12}$~G and an initial spin period $P_0 < 50$~ms.  These values
are consistent with ``typical'' radio pulsars such as the Crab ($B =
4\times10^{12}$~G, $P_0 = 19$~ms) and Vela ($B = 3\times10^{12}$~G)
pulsars, but not with the emerging class of young pulsars which are
highly magnetized and/or slow initial rotators, such as PSRs~J1846--0258
($B=5\times10^{13}$~G; \cite{gvbt00}) or J1210--5226 ($P_0 \approx
400$~ms; \cite{pzst02}).

\subsection{Evolutionary State of the PWN}

Chevalier (1998\nocite{che98}) divides PWN evolution into successive
phases.  The PWN first expands supersonically in the SNR interior.
The PWN then collides with the SNR reverse shock; this compresses the PWN
and causes it to expand more slowly.  The pulsar's motion can later become
supersonic in the shocked SNR interior, and it then drives a bow-shock.

If we assume that the pulsar
spin-down luminosity is constant in the initial supersonically expanding phase,
the radius of the PWN
evolves such that 
$R_{PWN} = 0.839 \times (\dot{E} \tau/E_0)^{1/5} V_0 \tau$,
where $V_0$ was the initial expansion velocity of the SNR
(\cite{vagt01}).
Here we observe $R_{PWN} \approx 3$~pc, $\tau \approx 13$~kyr
and $E_0 \approx 5\times10^{50}$~ergs, and we adopt
$\dot{E} = E_{37}\times10^{37}$~ergs~s$^{-1}$. Thus if the PWN is
still in this phase of evolution,
we require $V_0 \approx 700E_{37}^{-1/5}$~\kms,
implying an enormous
ejected mass $M_{ej} = 2E_0/V_0^2 \approx 100E_{37}^{2/5}~M_\odot$. 
We thus conclude that the PWN is almost certainly
not expanding supersonically, having 
a radius much smaller than expected in this interpretation.

The brightest radio and X-ray emission from the PWN is offset from the
SNR's geometric center by $l = (2.7\pm0.5)d_0$~pc. Since this peak
most likely marks the current position of the unseen
pulsar, a possible interpretation is that the pulsar is moving away
from its birth site with a transverse velocity $V_T =
l/\tau = 200\pm40$~\kms. The radio PWN is clearly elongated, with a
fading tail pointing back along the implied direction of motion. While
this is suggestive of a bow shock morphology,
this requires that the pulsar be moving supersonically through shocked
ejecta. In the Sedov solution, this condition is met only when $l/R_{SNR}
\ga 0.7$ (\cite{vag98}). Unless the pulsar's motion is at less than
$\sim20^\circ$ to the line-of-sight (implying a true velocity $V >
700$~\kms), the pulsar in this system is still too close to the center
of its SNR to meet this requirement.

The remaining possibility is that this PWN has undergone a reverse
shock interaction with its SNR, and is now expanding
subsonically. Such an interaction begins at an age
$\sim10$~kyr when $M_{sw} \ga 10~M_{ej}$
(\cite{che98}), consistent with the properties of this
source. The elongated morphology and comparatively small radius seen for
the radio PWN are simply accounted for by reverse-shock compression;
the same properties have been similarly interpreted in the other
SNRs listed in Table~\ref{tab_compare} (\cite{bcf01}).

\section{Conclusions}

New radio and X-ray observations of SNR~\snr\ in the LMC 
demonstrate this source to consist of both
an outer shell and a central pulsar-powered nebula.  We have used the
properties of this SNR to infer that the unseen central engine is
a typical young radio pulsar, with an initial period $P_0 < 50$~ms,
a current period $P \sim 100$~ms, a surface field $B < 6 \times
10^{12}$~G and a spin-down luminosity $\dot{E} \sim 10^{37}$~erg~s$^{-1}$.
The small and elongated radio nebula results from
compression of the PWN by the SNR reverse shock.

This study demonstrates that several useful constraints on the properties
of an unseen pulsar can be inferred from those of its associated PWN
and SNR, provided that good estimates for the system's distance and age
are available.  With many new studies of PWNe and composite SNRs now
emerging, such an approach can be applied to many other sources.

\begin{acknowledgements}

We thank Bob Sault for allowing us to carry out the ATCA observations
remotely from Harvard University, Dick Manchester for information on
pulsar surveys of the LMC, and Eric van der Swaluw and Patrick Slane for
useful discussions.  The Australia Telescope is funded by the Commonwealth
of Australia for operation as a National Facility managed by CSIRO.
This work was supported by NASA through grants GO1-2075X and NAG5-13032.

\end{acknowledgements}


\clearpage

\begin{table}[hbt]
\begin{center}
\caption{Radio properties of SNR~\snr.}
\label{tab_src}
\begin{tabular}{ccccccccccc} \hline
Waveband  & Resolution & \multicolumn{3}{c}{Flux Density (mJy)}   \\
(cm)      & (arcsec) & Core & Shell & Total  \\ \hline
1.4~GHz &$7.3\times6.7$ & $46\pm2$ & $140\pm2$ & $186\pm3$ \\
2.4~GHz & $9.2\times8.4$ & $46\pm2$ & $105\pm2$ & $151\pm3$ \\ \hline
\end{tabular}
\end{center}
\end{table}

\begin{table}
\begin{center}
\caption{Comparison of SNR~\snr\ to other similar systems.}
\label{tab_compare}
\begin{tabular}{lccccccc} \hline
SNR & Age         & $R_{SNR}$         &  $R_{PWN}$    &
$L_{R,PWN}$
& $L_{X,PWN}$  &
Ref\tablenotemark{b} \\
    &  ($10^3$~yr) & \multicolumn{2}{c}{(pc)}  & 
\multicolumn{2}{c}{($10^{34}$~ergs~s$^{-1}$)}  \\ \hline
\snr &  13 & $29\pm1$ &  $7\times5$ & 2 & 6 &  1 \\
Vela~SNR & $\sim11$ & 42 & 10 & 0.8 & 0.8\tablenotemark{a} & 2, 3 \\
G0.9+0.1 & $\sim10-20$ & 20 & 6 & 7 & 6 & 4, 5 \\ \hline
\end{tabular}
\tablenotetext{}{$R_{SNR}$ and $R_{PWN}$ are the radii of
the SNR and PWN respectively;
$L_{R,PWN}$ and $L_{X,PWN}$ are the approximate radio 
and X-ray luminosities, respectively, of the PWN. }
\tablenotetext{a}{The PWN luminosity quoted by Helfand \etal\
(2001\nocite{hgh01}) is much lower than this, but
corresponds to only the innermost component of the PWN.}
\tablenotetext{b}{{\sc References:} (1) This paper;
(2) Weiler \& Panagia (1980\nocite{wp80}); (3)
Helfand \etal\ (2001\nocite{hgh01}); (4) Helfand \& Becker
(1987\nocite{hb87}); (5) Porquet \etal\ (2003\nocite{pdw03}).}
\end{center}
\end{table}

\normalsize

\clearpage

\clearpage

\end{document}